\documentclass[10pt,prb]{revtex4}
\usepackage{epsfig}
\usepackage{amsmath}

\begin{document}

\title{Exact results on diffusion in a piecewise linear potential with a time dependent sink.}
\author{ Diwaker and Aniruddha Chakraborty \\
School of Basic Sciences, Indian Institute of Technology Mandi, Mandi, Himachal-Pradesh 175001, India.}

\begin{abstract}
The Smoluchowski equation with a time dependent sink term is solved exactly. In this method by knowing the probability distribution at the origin $P(0,s)$, one may derive the probability distribution at all positions {\it i.e.,} $P(x,s)$. Further the exact solution for Smoluchowski equation are also provided in different cases where the sink term has linear, constant, inverse and exponential variation in time.  
\end{abstract}

\maketitle
\section{Introduction}
\noindent 
The diffusion process of a particle in a potential having sink is studied through the solution of the Smoluchowski equation\cite{diw1}. It is of much interest to many scientists in chemical dynamics as it serves  a reference model for a wide variety of dynamical processes. A huge number of attempts has been made to study this diffusion processes with a suitable position of the sink \cite{diw2,diw3,diw23}. The reason being as it find many applications in diffusion controlled reactions\cite{diw4}. Such a model is used by Wilemski and Fixman\cite{diw5,diw6} to calculate the rate of diffusion controlled reactions as well as cyclization of polymer chain in solutions. Ovchinnikova\cite{diw7}, Zusman\cite{diw8}, Marucs and Nadler\cite{diw9} have used such a model for electron transfer reactions in polar solvents. Marcus\cite{diw10} recently used a diffusive equation with a sink term to develop a theory of uni-molecular reactions in clusters. Pressure influence on isomerization reactions is also explained by Sumi\cite{diw11} using such type of model. Bagchi, Fleming and Oxtoby\cite{diw12,diw13} uses a model of this kind to analyze barrier less electron relaxation in solution.   Exact analytical results for diffusion problems helps in understanding the different parameters like friction and provide an insight to different approximations. Most of these work had focussed on the time evolution/propagator derivation of the case of one or more Dirac Delta function sinks with constant strength in time. There has been a huge work on the analytical solution of Smoluchowski equation in context to various problems which do not involve any kind of time dependence or in other words there are no cases where the Smoluchowski equation with a time dependent sink is solved by analytical methods. Even there is no analytical solution available for the simplest possible case {\it i.e.,} for free particle. There are large number of works on diffusion under time independent sink  \cite{diw14}.  In contrast, this paper is devoted to the concept that deals mainly with a Dirac Delta sink whose strength varies with time and we are the first one to consider this effect explicitly. There are two common methods that one may think of for solving this problem. One method is using a path integral method of the Feynman type and the other is using Laplace transforms. We use the latter method though the two methods are closely related. Our method closely follows the method used for solving Schrodinger wave equation.

\section{Formulation of the problem}       

\noindent We would like to solve the Smoluchowski equation with a time-dependent sink for the problem as shown in Figure $1$.
\begin{figure}
\centering 
{\includegraphics[width=90mm]{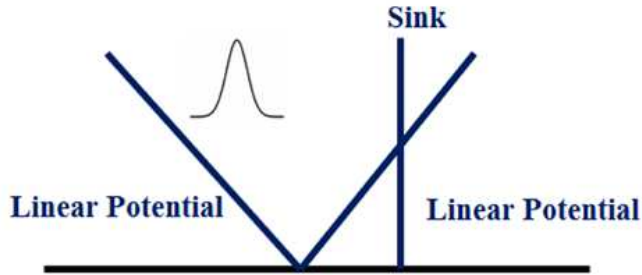}}\caption{
Schematic diagram showing the formulation of our problem}
\end{figure}
The simplest piecewise linear potential can be represented by the following equations i.e. 
\begin{equation}
\frac{\partial P(x,t)}{\partial t} = D\frac{\partial}{\partial x}\left[\frac{\partial P(x,t)}{\partial x}+P(x,t)\frac{\partial U(x)}{\partial x}\right]
\end{equation}
In our case we consider 
\begin{eqnarray}
U(x) = -x\omega,\; -\infty \leq x \leq 0 \nonumber \\
U(x) = x\omega,\; 0 \leq x \leq \infty \nonumber \\
U(x) = \omega |x|
\end{eqnarray}
In lieu of equation $(2)$ equation $(1)$ can be written as
\begin{equation}
\frac{\partial P(x,t)}{\partial t} = D\frac{\partial}{\partial x}\left[\frac{\partial P(x,t)}{\partial x}+P(x,t)\omega \right]
\end{equation}
Taking the Laplace transform of above equation we get as
\begin{equation}
S\overline{P}(x,s)-P(x,0) = D\left[\frac{\partial^2}{\partial x^2}\overline{P}(x,s)+\omega\frac{\partial}{\partial x}\overline{P}(x,s)\right]
\end{equation}
which can be further rewritten as
\begin{eqnarray}
-D\frac{\partial^2}{\partial x^2}\overline{P}(x,s)-\omega\frac{\partial}{\partial x}\overline{P}(x,s)+S\overline{P}(x,s) = P(x,0)\nonumber \\
\left[-D\frac{\partial^2}{\partial x^2}-\omega\frac{\partial}{\partial x}+S\right]\overline{P}(x,s) = P(x,0)
\end{eqnarray}
Integrating equation $(5)$ from $(x_{0}-\epsilon)$ to $(x_{0}-\epsilon)$ and using the notations $p^2 = q^2+\frac{S}{D}$, $S = D{(p^2-q^2)}$ and $q=\frac{\omega}{2}$, the general solution\cite{diw15,diw16} of the above equation in the two regions of interest can be written as 
\begin{equation}
\overline{P}(x,s) = a(s)e^{-(p+q)|x|}+\frac{1}{2D(p+q)}\int_{-\infty}^{\infty} e^{-(p+q)|x-x_{0}|} P(x_{0}) dx_{0}
\end{equation}
The above equation will determine the probability distribution at all positions. Further subsections will determine the variation of sink in different cases.
\section{ Introducing the Time dependent sink term into Smoluchowski equation and its exact solution }
We will consider the case such as
\begin{eqnarray}
U(x) = -x\omega,\; -\infty \leq x \leq 0 \nonumber \\
U(x) = x\omega,\; 0 \leq x \leq \infty \nonumber \\
U(x) = \omega |x|
\end{eqnarray}
The Smoluchowski equation with a time dependent sink term can be written as
\begin{equation}
\frac{\partial P(x,t)}{\partial t} = D\frac{\partial}{\partial x}\left[\frac{\partial P(x,t)}{\partial x}+P(x,t)\left(\frac{\partial U}{\partial x}\right)\right]+2k(t)\delta(x)P(x,t)
\end{equation}
for $0\leq x \leq \infty$ and $-\infty \leq x \leq \infty$, the Smoluchowski equation can be written as
\begin{eqnarray}
\frac{\partial P(x,t)}{\partial t} = D\frac{\partial^2}{\partial x^2}P(x,t)+\omega D\frac{\partial}{\partial x}P(x,t)+2k(t)\delta(x)P(x,t),\; for\;x>0\nonumber \\
\frac{\partial P(x,t)}{\partial t} = D\frac{\partial^2}{\partial x^2}P(x,t)-\omega D\frac{\partial}{\partial x}P(x,t)+2k(t)\delta(x)P(x,t),\; for\;x<0
\end{eqnarray}
we will now solve equation $(9)$, however we must consider the homogeneous equations in order to satisfy the boundary conditions at the origin i.e.
\begin{eqnarray}
\frac{\partial P(x,t)}{\partial t} = D\frac{\partial^2}{\partial x^2}P(x,t)+\omega D\frac{\partial}{\partial x}P(x,t)\nonumber \\
\frac{\partial P(x,t)}{\partial t} = D\frac{\partial^2}{\partial x^2}P(x,t)-\omega D\frac{\partial}{\partial x}P(x,t)
\end{eqnarray}
The solution of both these equations for all x, positive and negative may be read as
\begin{equation}
\overline{P}(x,s) = a(s)e^{-(p+q)|x|}+\frac{1}{2D(p+q)}\int_{-\infty}^{\infty} e^{-(p+q)|x-x_{0}|} P(x_{0}) dx_{0}
\end{equation}
further to determine the constant a(s), we will consider the Laplace transform of equation$(9)$ which can be written as
\begin{equation}
S\overline{P}(x,s)-P(x,0) = D\frac{\partial^2}{\partial x^2}\overline{P}(x,s)+\omega D \frac{\partial}{\partial x}\overline{P}(x,s)+2\delta(x)L\left[k(t)P(x,t)\right]
\end{equation}
Integrating the above equation from $x = 0-\epsilon$ to $0+\epsilon$ we get 
\begin{equation}
\left[\frac{\partial}{\partial x}\overline{P}(x,s)+\omega \overline{P}(x,s)\right]_{0-\epsilon}^{0-\epsilon}+\frac{2}{D}L\left[k(t)P(0,t)\right] = 0
\end{equation}
above equation again can be rewritten as
\begin{eqnarray}
\left[\frac{\partial \overline{P}(x,s)}{\partial x}\right]_{0-\epsilon}^{0+\epsilon}+2 \omega \overline{P}(0,s)+\frac{2}{D}L\left[k(t)P(0,t)\right] = 0 
\end{eqnarray}
further the above equation can be simplified as
\begin{equation}
a(s)\left[p+q-\omega \right] = \frac{L}{D}\left[k(t)P(0,t)\right]-\frac{\omega}{2D}\int dx_{0}e^{(p+q)x_{0}}P(x_{0})
\end{equation}
putting $P(x_{0}) = \delta(x-x_{0})$, the above equation can be further rewritten as
\begin{equation}
a(s) = \frac{L\left[k(t)P(0,t)\right]}{D \left[p+q-\omega \right]}-\frac{\omega}{2D\left[p+q-\omega \right](p+q)}e^{(p+q)|x_{0}|}
\end{equation}
using this value of constant a(s) in equation no $(11)$ will help us in determining the probability distribution in the case with time dependent sink i.e.
\begin{equation}
\overline{P}(x,s) =\frac{L\left[k(t)P(0,t)\right]}{D \left[p+q-\omega \right]}e^{(p+q)|x|}-\frac{\omega}{2D\left[p+q-\omega \right](p+q)}e^{(p+q)|x+x_{0}|}+\frac{1}{2D(p+q)}\int_{-\infty}^{\infty} e^{(p+q)|x-x_{0}|}P(x_{0})dx_{0}
\end{equation}
which finally will become
\begin{equation}
\overline{P}(x,s) =\frac{L\left[k(t)P(0,t)\right]}{D \left[p+q-\omega \right]}e^{(p+q)|x|}-\frac{\omega}{2D\left[p+q-\omega \right](p+q)}e^{(p+q)|x+x_{0}|}+\frac{1}{2D(p+q)}e^{(p+q)|x_{0}|}
\end{equation}
\section{Determination of Probability distribution \textsc{P{(x,t)}}, constant a(s) with time dependent sink term k(t) = {\large $\alpha_{0}$}}
In this case we will consider $k(t) = \alpha_{0}$, hence the sink term can be written as
\begin{equation}
L\left[k(t)P(0,t)\right] = \int_{0}^{\infty}e^{-st}\alpha_{0}P(0,t)dt = \alpha_{0}\overline{P}(0,s)
\end{equation}
the constant a(s) will become
\begin{equation}
a(s) = \frac{\left[\alpha_{0}\overline{P}(0,s)\right]}{D \left[p+q-\omega \right]}-\frac{\omega}{2D\left[p+q-\omega \right](p+q)}e^{(p+q)|x_{0}|}
\end{equation}
Putting this value of a(s) in equation $(10)$ we will get 
\begin{equation}
\overline{P}(x,s)=\frac{\alpha_{0}\overline{P}(0,s)e^{(p+q)|x|}}{D \left[p+q-\omega \right]}-\frac{\omega e^{(p+q)|x+x_{0}|}}{2D\left[p+q-\omega \right](p+q)}+\frac{1}{2D(p+q)}e^{(p+q)|x-x_{0}|}
\end{equation}
In above equation putting the value x = 0, we will find P(0,\;s) given as
\begin{equation}
\overline{P}(0,s) = e^{(p+q)|x_{0}|}\left[\frac{-2\omega+p+q}{2(p+q)(p+q-\omega-\alpha_{0})}\right]
\end{equation}
 putting this value of $\overline{P}(0,s)$ back into equation $(21)$and after some simplification we will get the probability distribution function as.
\begin{equation}
\overline{P}(x,s) = \frac{e^{(p+q)|x+x_{0}|}}{2D(p+q-\omega)(p+q)}\left[\frac{(p+q-\omega)\alpha_{0}}{D(p+q-\omega-\alpha_{0}}-\omega\right]+\frac{1}{2D(p+q)}e^{(p+q)|x-x_{0}|}
\end{equation}
\section{Solution of Smoluchowski equation with time dependent sink term k(t) = $-\alpha$\lowercase{t}}
In this case we will consider $k(t) =  -\alpha t$, hence the sink term can be written as
\begin{equation}
L\left[k(t)P(0,t)\right] = L\left[-\alpha t P(0,t)\right]= \int_{0}^{\infty}e^{-st}-\alpha t P(0,t)dt = \alpha \frac{\partial}{\partial s}\overline{P}(0,s)
\end{equation}
We will follow the same methodology as discussed in section (III) and will reach directly to the equation written as $(13)$ i.e.
\begin{eqnarray}
\left[\frac{\partial \overline{P}(x,s)}{\partial x}\right]_{0-\epsilon}^{0+\epsilon}+2 \omega \overline{P}(0,s)+\frac{2}{D}L\left[k(t)P(0,t)\right] = 0 
\end{eqnarray}
from the above equation we will find the value of a(s)which can  be written as
\begin{equation}
a(s) = \frac{1}{D(p+q)}\alpha\frac{\partial}{\partial s}\overline{P}(0,s)+\frac{\omega}{(p+q)}\overline{P}(0,s)
\end{equation}
putting this value of a(s) in equation $(10)$ we will get the probability distribution function as
\begin{equation}
 \overline{P}(x,s) = \frac{1}{D(p+q)}\alpha\frac{\partial}{\partial s}\overline{P}(0,s)e^{-(p+q)|x|}+\frac{\omega}{(p+q)}\overline{P}(0,s)e^{(p+q)|x|}+\frac{1}{2D(p+q)}e^{-(p+q)|x-x_{0}|}
\end{equation}
putting x =0 in the above equation and after simplification we will reach at following equation given below
\begin{equation}
\frac{\partial}{\partial s}\overline{P}(0,s)-\left(\frac{pD}{\alpha}-\frac{Dq}{\alpha}\right)\overline{P}(0,s) = -\frac{1}{2D}e^{-(p+q)|x_{0}|}
\end{equation}
multiplying the above equation on both sides by $e^{f(s)}$, where $f^{'}(s)=\frac{\partial}{\partial s}f(s) = \frac{D}{\alpha}(p-q)$ we can rewrite the above equation as
\begin{equation}
e^{f(s)}\frac{\partial}{\partial s}\overline{P}(0,s)-e^{f(s)}\left(\frac{pD}{\alpha}-\frac{Dq}{\alpha}\right)\overline{P}(0,s) = -\frac{e^{f(s)}}{2D}e^{-(p+q)|x_{0}|}
\end{equation}
using suitable integration and simple mathematics we will reach finally at
\begin{equation}
\overline{P}(0,s) = \frac{e^{-f(s)}}{2D}\int_{s}^{\infty}e^{f(s)}e^{-(p+q)|x_{0}|}
\end{equation}
where
\begin{equation}
f(s) =\left[ \frac{-2}{2\alpha (q^2+\frac{S}{D})^{\frac{3}{2}}}- \frac{DqS}{\alpha}\right]
\end{equation}
hence, using above equation we are able to determine $\overline{P}(0,s)$ and once it is known we are able to determine the probability distribution function $\overline{P}(x,s)$ everywhere
\section{Determination of Probability distribution \textsc{P{(x,t)}}, constant a(s) with time dependent sink term k(t) =\large{$\frac{\alpha}{\lowercase{t}}$}}
In this case we will consider $k(t) = \frac{\alpha}{t}$, hence the sink term can be written as
\begin{equation}
L\left[k(t)P(0,t)\right] = \int_{0}^{\infty}e^{-st}\frac{\alpha}{t}P(0,t)dt = \alpha\int_{s}^{\infty}ds^{'}\overline{P}(0,s)
\end{equation}
we will again consider equation $(13)$ written as
\begin{eqnarray}
\left[\frac{\partial \overline{P}(x,s)}{\partial x}\right]_{0-\epsilon}^{0+\epsilon}+2 \omega \overline{P}(0,s)+\frac{2}{D}L\left[k(t)P(0,t)\right] = 0 
\end{eqnarray}
from the above equation we will determine the constant a(s)which can be written as
\begin{equation}
a(s) = \frac{\omega\overline{P}(0,s)}{D \left[p+q\right]}+\frac{\alpha}{D\left[p+q\right]}\int_{s}^{\infty}ds^{'}\overline{P}(0,s)
\end{equation}
Putting this value of a(s) in equation $(10)$ we will get 
\begin{equation}
\overline{P}(x,s)=\frac{\omega\overline{P}(0,s)e^{-(p+q)|x|}}{D \left[p+q\right]}+\frac{\alpha e^{-(p+q)|x|}}{D\left[p+q\right]}\int_{s}^{\infty}ds^{'}\overline{P}(0,s)+\frac{1}{2D(p+q)}e^{-(p+q)|x-x_{0}|}
\end{equation}
using x = 0 and some simple modification in above equation we can rewrite it as
\begin{equation}
\overline{P}(0,s) = \frac{\alpha}{Dp}\int_{s}^{\infty}ds^{'}\overline{P}(0,s)-\frac{\omega}{2Dp}[D+2]\overline{P}(0,s)+\frac{1}{2Dp}e^{-(p+q)}|x_{0}|
\end{equation}
in solving the above equation we will consider $u(s) = \int_{s}^{\infty}ds^{'}\overline{P}(0,s)$, hence the above equation can be further modified as
\begin{equation}
\frac{d u(s)}{ds}-\frac{2\alpha}{[2Dp+\omega p+2\omega]}u(s)= \frac{2}{[2Dp+\omega p+2\omega]}e^{-(p+q)|x_{0}|}
\end{equation}
multiplying the above equation on both sides by $e^{f(s)}$, where $f^{'}(s)=\frac{\partial}{\partial s}f(s) = \frac{2\alpha}{[2Dp+\omega p+2\omega]}$ we can rewrite the above equation as
\begin{equation}
e^{f(s)}\frac{d u(s)}{ds}-e^{f(s)}\frac{2\alpha}{[2Dp+\omega p+2\omega]}u(s)= \frac{2e^{f(s)}}{[2Dp+\omega p+2\omega]}e^{-(p+q)|x_{0}|}
\end{equation}
using integration and suitable mathematics for simplification we can rewrite the above equation as
\begin{equation}
u(s) = e^{-f(s)}\int_{s}^{\infty}\frac{e^{f(s)}}{[2Dp+\omega p+2\omega]}e^{-(p+q)|x_{0}|}
\end{equation}
with $f(s) = \int_{0}^{s}\frac{2\alpha}{2D(q^2+\frac{s}{D})^{\frac{1}{2}}+2\omega+2D}ds$
we can revert back and calculate $\overline{P}(0,s)$ by using the expression $u(s) = \int_{s}^{\infty}ds^{'}\overline{P}(0,s)$. Once we know $\overline{P}(0,s)$, we can calculate the probability distribution function $\overline{P}(x,s)$ everywhere. 
\section{Solution of Smoluchowski equation with time dependent sink term k(t) = $\beta$ exp(-$\alpha\; t$)}
In this case we will consider k(t) = $\beta$ exp(-$\alpha\; t$), hence the sink term can be written as
\begin{equation}
L\left[k(t)P(0,t)\right] = L\left[\beta exp (-\alpha t)P(0,t)\right]= \int_{0}^{\infty}e^{-st}\beta exp (-\alpha t) P(0,t)dt =\beta \overline{P}(0,s+\alpha)
\end{equation}
again using equation $(13)$ 
\begin{eqnarray}
\left[\frac{\partial \overline{P}(x,s)}{\partial x}\right]_{0-\epsilon}^{0+\epsilon}+2 \omega \overline{P}(0,s)+\frac{2}{D}L\left[k(t)P(0,t)\right] = 0 
\end{eqnarray}
we will evaluate the value of a(s) i.e.
\begin{equation}
a(s) = \frac{\omega}{(p+q)}\overline{P}(0,s)+\frac{\beta}{D(p+q)}\overline{P}(0,s+\alpha)
\end{equation} 
Putting this value of a(s) in equation $(10)$ we will get the probability distribution function as
\begin{equation}
 \overline{P}(x,s) = \frac{\omega}{(p+q)}\overline{P}(0,s)e^{-(p+q)|x|}+\frac{\beta}{D(p+q)}\overline{P}(0,s+\alpha)e^{-(p+q)|x|}+\frac{1}{2D(p+q)}e^{-(p+q)|x_{0}|}
\end{equation}
Putting x = 0 in the above equation and using some simplification the above equation can be rewritten as
\begin{equation}
\overline{P}(0,s) = \frac{1}{2Dp-2\omega}e^{-(p+q)|x_{0}|}+\frac{2\beta}{2Dp-2\omega}\overline{P}(0,s+\alpha)
\end{equation}
further to obtain a solution we could iterate the expression repeatedly to obtain the series solution as
\begin{equation}
\overline{P}(0,s) = \sum_{n=0}^{\infty}\gamma^{n}\tau_{n}(s)
\end{equation}
using $(45)$ in $(44)$ and solving for $\tau's$ by equating like powers of $\gamma$ we find that
\begin{eqnarray}
\tau_{n}(s) = \frac{1}{2D(q^2+\frac{s}{D})^{\frac{1}{2}}-2\omega}e^{-((q^2+\frac{s}{D})^{\frac{1}{2}}+q)|x_{0}|}\; for\; n = 0 \nonumber \\
\tau_{n}(s) =\Pi_{j = 0}^{n-1} \frac{(2)^{n}e^{-((q^2+\frac{s}{D})^{\frac{1}{2}}+q)|x_{0}|}}{2D(q^2+\frac{(s+j\alpha)}{D})^{\frac{1}{2}}-2\omega}\; for \; n>0
\end{eqnarray}
\section{Conclusions}
In the current work the authors had found the exact solution for the Smoluchowski equation with a time dependent delta function sink in many special cases. These special cases includes constant, linear, inversely  and exponential time dependence. To conclude, the treatment of the time dependent sink term is not reported till now and we are the first to provide its analytical treatment explicitly.

\section{Acknowledgements}
Th authors wants to thank IIT Mandi for Professional Development Fund as well as HTRA scholarship.

\end{document}